\documentstyle[prl,multicol,aps,epsfig]{revtex}

\renewcommand{\narrowtext}{\begin{multicols}{2} \global\columnwidth20.5pc}
\renewcommand{\widetext}{\end{multicols} \global\columnwidth42.5pc}

\begin{document}
\bibliographystyle{prsty}
\title{Universal Features of Interacting Chaotic Quantum Dots.\\
Application to Statistics of Coulomb Blockade Peak Spacings. }
\author{Victor Belinicher$^{(1,2)}$, Eran Ginossar$^{(1)}$, 
and Shimon Levit$^{(1)}$}
\address{(1) Weizmann Institute of Science, Rehovot 76100, Israel \\ 
(2)Weston Visiting Professor, Weizmann Institute of Science; \\
Institute of Semiconductor Physics, 630090, Novosibirsk, Russia}

\maketitle
\begin{abstract}
We  present a complete classification of the electron-electron interaction in
chaotic quantum dots based on expansion in  inverse powers of $1/M$, 
the number of the electron states in the Thouless window, $M \simeq k_F R$. 
This classification is quite universal and extends and
enlarges the universal non interacting RMT statistical ensembles.  We show 
that existing Coulomb blockade peak spacing data for $B=0$ and $B\ne 0$
is described quite
accurately by the interacting GSE and by its extension to $B\ne 0$. The bimodal structure existing in the
 interacting 
GUE case is completely washed out by the combined effect of the spin orbit,
 pairing and higher order residual interactions. 
\end{abstract}

\date{\today}
\maketitle
\pacs{73.23.Hk, 73.63.Kv}
\narrowtext
Measurement of the fluctuations of the Coulomb blockade (CB) peak spacings
as well as the mesoscopic spin effects provide an excellent probe of the
properties of interaction effects in disordered quantum dots (QD). 
Basic experiments
where performed \cite{Siv1,Fol1} in ballistic, well isolated QD of
irregular shape formed in $2d$ GaAs heterostructures. The QD sizes were $%
R\gg 1/k_{F}$, where $k_{F}$ is the electron Fermi wave vector.

The main general theme underlying this problem is an interplay between chaos
and interactions in quantum mechanical motion of electrons in a restricted
geometry. In a recent work \cite{Alt1},
cf., also \cite{Hal1}, on mesoscopic spin effects a proposal appeared for a
universal Hamiltonian which controls the main physics of interactions in a
chaotic QD in the extreme limit $k_{F}R\gg 1$. In \cite{Bar1} statistical
fluctuations around this limit were included in the framework of the
Hartree-Fock (HF) method. Our main goal here is to show that
these fluctuations and indeed the entire interacting Hamiltonian can be
represented and {\em classified }in a unifying {\em \ universal }scheme, cf.
Eqs. (\ref{9},\ref{9a},\ref{9b}) below, which follows and extends the universal
symmetry classes of the Wigner-Dyson statistical theory for non interacting
electrons. Our second goal is to apply this theory to the problem of
fluctuations of the CB peak spacings. Although this has received much
attention recently \cite{BlMM,Berk1,Lev1,Wal1,Bar1}, a
consistent description is still lacking. We will show that the universal
interacting GSE Hamiltonian and its extension (we call it GUSE) to the non zero
 magnetic field in our scheme accounts quite
well for the experimental distributions, cf. Fig. 1 below. The GSE choice
matches perfectly the recently discovered strong spin-orbit (SO) effects in
GaAs QD, \cite{Hal1}

The Hamiltonian of an interacting QD consists of one and two body parts, 
$H=H_{0}+H_{int}$.
\begin{eqnarray} \label{1}
&&H_{0}=\sum_{a,b,\sigma ,\sigma ^{\prime }}H_{a\sigma ,b\sigma ^{\prime
}}a_{a\sigma }^{\dagger }a_{b\sigma ^{\prime }}=\sum_{\alpha \nu }\epsilon
_{\alpha }a_{\alpha \nu }^{\dagger }a_{\alpha \nu },
\\ 
&&H_{int}=\frac{1}{2}\sum V_{\alpha \nu_1 ,\beta \nu_2;\gamma
\nu_3,\delta \nu_4 }a_{\alpha \nu_1 }^{\dagger }a_{\beta
\nu_2}^{\dagger }a_{\delta \nu_4}a_{\gamma \nu_3},
\end{eqnarray}
Here indices $a,b$ denote space orbitals while $\sigma ,\sigma ^{\prime }$
are the spin indices. With a possible SO interaction 
$H_{a\sigma ,b\sigma ^{\prime }}$ is in general not diagonal in the spin
indices. We use  $\alpha ,\nu $ to numerate the eigen states of 
$H_{a\sigma ,b\sigma ^{\prime }}$; $\epsilon _{\alpha }$ are its eigen energies
 with $\alpha $ the orbital  and $\nu=\pm \frac{1}{2}$ - the spin or 
Kramers index in the presence of SO. In irregular QD the one
electron Hamiltonian in the Thouless window of states  can be described
 by a random matrix theory (RMT),\cite{Met1,Been1,Alh1}. We will denote by $M$ the rank of 
$H_{a\sigma ,b\sigma ^{\prime }}$. 
The statistics of  $H_{a\sigma ,b\sigma ^{\prime }}$ depends on
the symmetry of the problem classified by standard RMT
ensembles - GOE, GUE and GSE.
The correlators of the eigen functions of $H_{a\sigma ,b\sigma ^{\prime }}$
depend on the ensemble.  For GOE they are
\begin{eqnarray}
&&<\psi _{\alpha \nu }^{\ast }({\bf r,\sigma })\psi _{\beta \nu ^{\prime }}
({\bf r}^{\prime },\sigma ^{\prime })>=<\psi _{\alpha \nu }({\bf r,\sigma }%
)\psi _{\beta \nu ^{\prime }}({\bf r}^{\prime },\sigma ^{\prime })> 
\\ \nonumber 
&&=<\psi
_{\alpha \nu }^{\ast }({\bf r,\sigma })\psi _{\beta \nu ^{\prime }}^{\ast }
({\bf r}^{\prime },\sigma ^{\prime })>=\delta _{\nu \sigma }\delta _{\nu
^{\prime }\sigma ^{\prime }}\delta _{\alpha \beta }K({\bf r},{\bf r}^{\prime
}).
\end{eqnarray}
We find it convenient to work in the coordinate-spin representation (${\bf r}
,\sigma )$.  
The function $K({\bf r},{\bf r}^{\prime })$ is 
\begin{eqnarray} \label{K}
&&K({\bf r},{\bf r}^{\prime })=\frac{1}{2M}\sum_{\alpha=1}^{M}\sum_{\nu \sigma }\psi
_{\alpha \nu }^{\ast }({\bf r,\sigma })\psi _{\alpha \nu }({\bf r}^{\prime
},\sigma)\\ \nonumber &&\simeq A^{-1}J_{0}(k_{F}|{\bf r}-
{\bf r}^{\prime }|).
\end{eqnarray}
Here $A$ is the area of QD and $J_{0}(x)$ is the zero order Bessel function
giving an approximate quasiclassical expression for $K({\bf r},{\bf r}^{\prime })$, 
\cite{Ber1}. For the GUE  the correlators of the type $<\psi \psi >$,
and $<\psi ^{\ast }\psi ^{\ast }>$ are zero while $<\psi ^{\ast }\psi >$ 
is the same as in GOE. For the GSE symmetry one has
\begin{eqnarray}  \label{7} 
&<&\psi _{\alpha \nu }^{\ast }({\bf r,\sigma })\psi_{\beta \nu ^{\prime }}(
{\bf r}^{\prime },\sigma ^{\prime })>=\frac{1}{2}\delta_{\nu \nu ^{\prime
}}\delta_{\sigma \sigma ^{\prime }}\delta_{\alpha \beta }K({\bf r},{\bf r}
^{\prime }), \\
&<&\psi _{\alpha \nu }({\bf r,\sigma })\psi _{\beta \nu ^{\prime }}({\bf r}
^{\prime },\sigma ^{\prime })>=\frac{1}{2}k_{\nu \nu ^{\prime }}k_{\sigma
\sigma ^{\prime }}\delta _{\alpha \beta }K({\bf r},{\bf r}^{\prime }),
\nonumber
\end{eqnarray}
$\hat{k}$ is $2\times 2$ time inversion matrix for spin $\frac{1}{2}$
systems \cite{Met1}. 

It is convenient to consider the interaction part of the Hamiltonian 
$H_{int}$ in the basis of the eigen functions $\psi_{\alpha \nu}$ of the
one electron Hamiltonian
\begin{eqnarray}\label{2}
&&V_{\alpha \nu_1 ,\beta \nu_2;\gamma \nu_3,\delta \nu_4}= 
V^*_{\gamma \nu_3,\delta \nu_4;\alpha \nu_1 ,\beta \nu_2}=
\int d{\bf r}d{\bf r}^{\prime}\times
\\ \nonumber 
&& \psi _{\alpha \nu_1}^{\ast}({\bf r},\sigma )
\psi_{\beta \nu_2}^{\ast}({\bf r}^{\prime},\sigma^{\prime})
U({\bf r},{\bf r}^{\prime})\psi _{\gamma \nu_3}({\bf r},\sigma)
\psi_{\delta \nu_4}({\bf r}^{\prime},\sigma^{\prime}).
\end{eqnarray}
Here $U({\bf r},{\bf r}^{\prime })$ is the the screened electron - electron
interaction in QD. If SO interaction is absent the space and spin coordinates
 ${\bf r}$ and $\sigma $ are separated.

It is possible to represent the interaction (\ref{2}) as a sum of parts
of different order in a small parameter $1/M$. An essential step in this
direction was made in \cite{Alt1}. Here we shall present the complete 
classification and investigate some of its consequences. We will only present
our main results
deferring the detailed derivations to Ref. \cite{BGL1}. It will be sufficient
here to assume that higher
correlators of $\psi $'s obey the rules of the Gaussian statistics. The role of 
the non Gaussian  corrections will be discussed in \cite{BGL1}. 
We use a cluster
decomposition of the matrix elements (\ref{2}) as fourth order polynomial
functions of the random wave functions $\psi $. Inserting this into $%
H_{int}$ we find that it consists of three groups of terms $%
H_{int}=H_{int}^{(0)}+H_{int}^{(1)}+H_{int}^{(2)}$. For the GOE these terms
are
\begin{eqnarray}\label{9} 
&&H_{int}^{(0)}=\frac{1}{2}V_{c}\hat{N}(\hat{N}-1)-J\hat{{\bf S}}^{2}+P
\hat{T}^{\dagger }\hat{T},  \\
&&H_{int}^{(1)}=\hat{N}\sum_{\alpha \beta }u_{\alpha \beta }^{n}(a_{\alpha
}^{\dagger }\cdot a_{\beta })+\hat{{\bf S}}\cdot \sum_{\alpha \beta
}u_{\alpha \beta }^{s}(a_{\alpha }^{\dagger }\mbox{\boldmath{$\sigma$}}
a_{\beta })+  \nonumber \\
&&+\hat{T}^{\dagger }\sum_{\alpha \beta }u_{\alpha \beta }^{p}a_{\alpha
\downarrow }a_{\beta \uparrow }-\sum_{\alpha \beta }u_{\alpha \beta }^{p\ast
}a_{\alpha \downarrow }^{\dagger }a_{\beta \uparrow }^{\dagger }\hat{T},
\nonumber \\
&&H_{int}^{(2)}=\frac{1}{2}\sum_{\alpha \beta \gamma \delta \sigma \sigma
^{\prime }}\tilde{V}_{\alpha \beta; \gamma \delta }a_{\alpha \sigma
}^{\dagger }a_{\beta \sigma ^{\prime }}^{\dagger }
a_{\delta \sigma^{\prime} }a_{\gamma \sigma}.  \nonumber
\end{eqnarray}
Here we denoted by dot the scalar product in the spin variables, $\hat{N}
=\sum_{\alpha }(a_{\alpha }^{\dagger }\cdot a_{\alpha })$ is the operator of
the total number of electrons, $\hat{{\bf S}}=\frac{1}{2}\sum_{\alpha
}(a_{\alpha }^{\dagger }\mbox{\boldmath{$\sigma$}}a_{\alpha })$ is the
operator of the total spin, $\mbox{\boldmath{$\sigma$}}$ are the Pauli
matrices and $\hat{T}=\sum_{\alpha }a_{\alpha \downarrow }a_{\alpha \uparrow
}$ is the total pairing operator. The decomposition (\ref{9}) is an identity which
is useful since as we will show below and in \cite{BGL1} it allows to
classify the groups of terms $H_{int}^{(0)}$, $H_{int}^{(1)}$ and 
$H_{int}^{(2)}$ by their degree of smallness with respect to $1/M.$ Namely,
apart of the capacitance term $V_{c}\sim M\Delta $ ($\Delta$ is mean level 
spacing)  and up to
logarithmic corrections the matrix elements in $H_{int}^{(n)}$ are $\sim
\Delta /(M)^{n/2}.$ The properties of $u$ and $\tilde{V}$ are discussed
below,cf. also Ref. \cite{BlMM,Bar1}.

For GUE the expression above remains valid except  that the terms
containing the pairing operators $T$ and $T^{\dagger }$ are absent. For GSE
the expression (\ref{9}) becomes
\begin{eqnarray}
&&H_{int}^{(0)}=\frac{1}{2}V_{c}\hat{N}(\hat{N}-1)+P\hat{T}^{\dagger }\hat{T}
,  \label{9a} \\
&&H_{int}^{(1)}=\hat{N}\sum_{\alpha \nu \beta \nu ^{\prime }}u_{\alpha \nu
\beta \nu ^{\prime }}^{n}a_{\alpha \nu }^{\dagger }a_{\beta \nu ^{\prime }}+
\nonumber \\
&&+\hat{T}^{\dagger }\sum_{\alpha \nu \beta \nu ^{\prime }}u_{\alpha \nu
\beta \nu ^{\prime }}^{p}a_{\alpha \nu }a_{\beta \nu ^{\prime
}}-\sum_{\alpha \nu \beta \nu ^{\prime }}u_{\alpha \nu \beta \nu ^{\prime
}}^{p\ast }a_{\alpha \nu }^{\dagger }a_{\beta \nu ^{\prime }}^{\dagger }\hat{%
T},  \nonumber \\
&&H_{int}^{(2)}=\frac{1}{2}\sum_{\alpha \nu_{1}\beta \nu_{2}\gamma \nu_{3}
\delta \nu_{4}}\tilde{V}_{\alpha \nu_{1}\beta \nu_{2};\gamma \nu_{3}
\delta \nu_{4}}
a_{\alpha \nu_1}^{\dagger }a_{\beta \nu_2}^{\dagger }
a_{\delta \nu_4}a_{\gamma \nu_3}.  \nonumber
\end{eqnarray}
Now the terms containing the spin operators disappear and the spin index
is replaced by the Kramers degeneracy index which we denote by $\nu$. The 
matrices $u^{n,p}$ and $\tilde{V}$ depend on $\nu$. 

Introduction of a perpendicular magnetic field removes the Kramers degeneracy
 and changes the statistics of the 
single particle Hamiltonian into GUE(2M). But this is not the interacting 
GUE obtained from (\ref{9}). The  second
correlator in  (\ref{7}) vanishes and consequently the  
terms containing the pairing operators $T$, $T^{\dagger }$ in (\ref{9a})
disappear. Also there is no need anymore for the Kramers spinor indices in
$u^n$ and $\tilde{V}$. Thus one obtains a different 1/M expansion which 
we term GUSE (unitary arising from simplectic) 
\begin{eqnarray} \label{9b}
&&H_{int}^{(GUSE)}=\frac{1}{2}V_{c}\hat{N}(\hat{N}-1)+
\hat{N}\sum_{\alpha,\beta=1}^{2M}u_{\alpha\beta}^{n}
a_{\alpha }^{\dagger }a_{\beta}+ \\
&&+\frac{1}{2}\sum_{\alpha,\beta,\gamma,\delta=1}^{2M}
\tilde{V}_{\alpha\beta;\gamma\delta}a_{\alpha}^{\dagger}
a_{\beta}^{\dagger}a_{\delta}a_{\gamma} \nonumber.
\end{eqnarray}
 Such an ensemble was first discussed in \cite{Hal1}. 

The lowest order $H_{int}^{(0)}$ in (\ref{9})  and its universality was fully
 discussed in \cite{Alt1}. The $\frac{1}{2}V_{c}\hat{N}(\hat{N}-1)$ term is at
the basis of the simplest Coulomb blockade theory while $J\hat{{\bf S}}^{2}$
appeared in relation to mesoscopic spin fluctuations,  \cite{Hal1}. Here we will
 discuss higher order terms and will then
explore their effect and interplay  with the $T^{\dagger }T$ term. We 
note, \cite{BGL1},  that $u$ and $\tilde{V}$ are such that 
$<u_{\alpha \beta }^{i}>$=\\ $<\tilde{
V}_{\alpha \beta \gamma \delta }>=0$ for $i=n,s,p$, and the average of the
second functional derivatives of $\tilde{V}$ with respect to $\psi $ or $%
\psi ^{\ast }$ is equal to zero. Up to corrections of the order $1/M$ the
matrices $u_{\alpha \beta }^{i},\ u_{\alpha \nu \beta \nu ^{\prime }}^{i}$
for $i=n,s,p$, are Gaussian random variables with zero average. For GOE one
finds, \cite{Bar1,BGL1} ,
\begin{eqnarray}
&&<u_{\alpha \beta }^{i\ast }u_{\alpha ^{\prime }\beta ^{\prime }}^{i^{\prime
}}>= \\ \nonumber &&=C_{ii^{\prime }}\frac{\Delta ^{2}}{M}\left( \delta _{\alpha \alpha
^{\prime }}\delta _{\beta \beta ^{\prime }}+\delta _{\alpha \beta ^{\prime
}}\delta _{\beta \alpha ^{\prime }}-\frac{2}{M}\delta _{\alpha \beta }\delta
_{\alpha ^{\prime }\beta ^{\prime }}\right) .
\end{eqnarray}
Here $C_{ii^{\prime }}=C_{i^{\prime }i}$ are 6 dimensionless constants which
depend on the average geometry of as well as on the details of the 
electron screening in QD (cf., below). On the basis of the correlators (\ref{7}) one
can find similar averages for GUE, GSE and GUSE. For GUE the correlators of 
the residual interaction $\tilde{V}$ in (\ref{9a}) are, \cite{Bar1,BGL1} 
\begin{eqnarray} \label{22} \nonumber
&&<\tilde{V}_{\alpha \beta; \gamma \delta }^{\ast }\tilde{V}_{\alpha ^{\prime}
\beta ^{\prime};\gamma ^{\prime }\delta ^{\prime }}>=
\frac{\Delta^{2}\ln M}{M^{2}}\times
\\ && \times D\lbrace \delta _{\alpha \alpha ^{\prime }}\delta _{\beta
\beta^{\prime }}\delta _{\gamma \gamma ^{\prime }}\delta _{\delta \delta
^{\prime }}+\delta _{\alpha \beta ^{\prime }}\delta _{\beta \alpha ^{\prime }}
\delta_{\gamma\delta^{\prime}}\delta_{\delta \gamma^{\prime}}  
+\\ \nonumber
&&+\delta _{\alpha \alpha ^{\prime }}\delta _{\beta
\beta ^{\prime }}\delta _{\gamma \delta ^{\prime }}\delta _{\delta \gamma
^{\prime }}+\delta _{\alpha \beta ^{\prime }}\delta _{\beta \alpha ^{\prime
}}\delta _{\gamma \gamma ^{\prime }}\delta _{\delta \delta ^{\prime}}
\rbrace +O(M^{-2}) 
\end{eqnarray}
where $D$ is again a dimensionless largely universal constant
(cf., below). One can easily write corresponding expressions for other
ensembles.

In order to proceed it is convenient to adopt the following decomposition of
the basic screened e-e interaction $U({\bf r},{\bf r})$ in (\ref{2}), cf., 
\cite{BlMM}
\begin{equation}
U({\bf r},{\bf r}^{\prime })=V_{c}+u^{sur}({\bf r})+u^{sur}({\bf r}^{\prime
})+V({\bf r},{\bf r}^{\prime }).
\end{equation}
Here $V_{c}$ is the constant capacitance part, cf., (\ref{9},\ref{9a}), $%
u^{sur}({\bf r})$ is the surface part of the potential caused by the
screening charges which are on the surface and $V({\bf r},{\bf r}^{\prime })$
is the screened bulk e-e interaction. One can express the constants entering in the
expressions (\ref{9},\ref{9a},\ref{9b}) in terms of $u^{sur}({\bf r})$ and $V({\bf r},%
{\bf r}^{\prime })$. For GOE one finds (expressions for other ensembles are
very similar), \cite{BGL1}
\begin{eqnarray} \label{23}
&&J=P=I_2, \ \ I_n=A^{n-2}\int K^{n}({\bf r},{\bf r}^{\prime })
V({\bf r},{\bf r}^{\prime })
d{\bf r}d{\bf r}^{\prime };   \\
&&u_{\alpha \beta }^{i}=\int u^{i}({\bf r},{\bf r}^{\prime })\psi _{\alpha
}^{\ast }({\bf r})\psi _{\beta }({\bf r^{\prime }})d{\bf r}d{\bf r}^{\prime
},i=n,s,p;  \nonumber
\end{eqnarray}
Here $u^{n}({\bf r},{\bf r}^{\prime })=\delta ({\bf r}-{\bf r}^{\prime })
(u^{sur}({\bf r})-\bar{u})-
(1/2)u^{s}({\bf r},{\bf r}^{\prime })$, $\ u^{p}({\bf r},{\bf r}
^{\prime })=u^{s}({\bf r},{\bf r}^{\prime })$ , $\bar{u}=A^{-1}\int u^{sur}
({\bf r})d{\bf r}$ and $u^{s}({\bf r},{\bf r%
}^{\prime })=K({\bf r},{\bf r}^{\prime })V({\bf r},{\bf r}^{\prime
})-J\delta ({\bf r}-{\bf r}^{\prime }).$ The matrix 
$\tilde{V}_{\alpha \beta;\gamma \delta }$ 
is given by (\ref{2}) if we substitute $U({\bf r},{\bf r}
^{\prime })\Rightarrow V({\bf r},{\bf r}^{\prime })$ and extract the irreducible
part. We also have
\begin{eqnarray} \label{24}
&&C_{ii^{\prime }}=\frac{M}{\Delta ^{2}}\int K(1,3)K(2,4)u^{i}(1,2)u^{i^{\prime }}(3,4)d\Omega,
 \\
&&D=\frac{M^2}{\Delta^2 \ln M}\int K^{2}(1,3)K^{2}(2,4)V(1,2)V(3,4)d\Omega,  \nonumber 
\end{eqnarray}%
where $V(1,2)\equiv V({\bf r}_{1},{\bf r}_{2})$, etc., and $d\Omega\equiv d%
{\bf r}_{1}d{\bf r}_{2}d{\bf r}_{3}d{\bf r}_{4}$.

We now turn to the problem of the fluctuations of the spacings
between Coulomb blockade peaks. We focus on the experiments in 2D GaAs dots,
Ref. \cite{Siv1}. An important observation made in \cite{Hal1}
was that the basic non interacting Hamiltonian for such dots must include a
strong SO interaction, the so called Rashba term \cite{Ras1}, $\alpha _{SO}
\hat{{\bf p}}\cdot \lbrack \hat{{\bf s}}\times {\bf n}],$ where 
$\hat{\bf p}$ and 
$\hat{\bf s}$ are the momentum and spin operators, ${\bf n}$ is the vector of the 
normal to the QD plane. The strength of this term in a typical GaAs/GaAlAs
 heterostructure is $\alpha _{SO}\simeq 2.5\cdot 10^{-7}mev\ cm/\hbar $ 
 \cite{Ras1}. The corresponding energy scale $\sim 0.3$ mev is $\gg \Delta$.
  Thus it is
appropriate to use the GSE ensemble for the random single electron
Hamiltonian and the expression (\ref{9a}) for the interaction. The
pairing term $T^{\dagger}T$ unlike other zeroth order terms {\em does not
commute} with the random single
electron part and should therefore {\em increase the effect} of fluctuations. 
Experiments in \cite{Siv1} included also the situation with an applied weak
perpendicular magnetic field. This corresponds to the GUSE Eq. (\ref{9b}). 

For the QD parameters we use $\Delta=2\hbar ^{2}/(m^{\ast }R^{2})$,
$E_{F}=\pi \hbar ^{2}/(m^{\ast }r_{sc}^{2})$,
 $r_{sc}\simeq n_{c}^{-1/2}$, where  $m^{\ast }$ is the effective mass, $n_c$ is the 
electron concentration in a QD. The Thouless energy is 
$E_{Th}=\hbar /\tau _{bal}=\sqrt{2\pi }\hbar^{2}/(m^{\ast }r_{sc}R)$.
  From Eq. (\ref{K}) it follows that the rank of RMT is $M\simeq \pi Rk_{F}/2$
so that  $M\simeq \pi ^{3/2}R/(2r_{sc})$. 
The  constants in the interacting part of the GSE Hamiltonian (\ref{9a})
  are completely determined by $u^{sur}({\bf r})$, $V({\bf r},{\bf r}^{\prime })$ and 
$K({\bf r},{\bf r}^{\prime })$, Eq.(\ref{K}). We take 
$u^{sur}({\bf r})=-(e^2/\epsilon^*\kappa R^2)(1-r^2/R^2)^{-1/2}$
which is appropriate for a 2D disc of radius R in the limit $\kappa R \gg 1$
where  $\kappa=r_{sc}^{-1}$.   For a disk
shape one gets an estimate   $C=2R\epsilon^*/\pi$, 
$V_c=e^2/C=M\Delta/(2\sqrt{2\pi})$. The screened interaction 
$V({\bf r},{\bf r}^{\prime })$ must behave as $e^2/\epsilon r$ for
$\kappa r \ll 1$ and as $e^2/\epsilon\kappa^2 r^3$ for $\kappa r \gg 1$, 
$r=|{\bf r}-{\bf r}^{\prime }|$. 
The constants P, 
C and D in Eqs.(\ref{23},\ref{24}) can be expressed \cite{BGL1} in terms of the 
integrals $I_n$ (\ref{23}), for $n=0,1,2$. They are sensitive to the 
intermediate range behavior of $V$. We estimated them as
$I_0=1.5\Delta, \ I_1=0.51\Delta, \ I_2=0.37\Delta$.

We treated the last term in (\ref{9},\ref{9a},\ref{9b}) in the
Hartree-Fock approximation. This and the term 
$\hat{N}\sum u_{\alpha \nu
\beta \nu ^{\prime }}^{n}a_{\alpha \nu }^{\dagger }a_{\beta \nu ^{\prime }}$
in (\ref{9a}) lead to a modified single particle part of H 
\centerline{$H^{HF}_{ab}=H_{ab}+Nu^{n}_{ab}+
\sum_{cd}\tilde{V}^A_{abcd}\rho_{dc}$,} where $\rho$ is one particle density
matrix, $\tilde{V}^A_{abcd}=\tilde{V}_{acdb}-\tilde{V}_{acbd}$ and
we omitted the spin indices. One can show, \cite{BGL1}, that statistical 
properties of the HF eigen values  and of the corresponding
eigen functions are practically the same as in the original RMT. 
The only noticeable effects appear when the particle-hole energy differences
are of order $\Delta/M$, \cite{Lev1,BGL1}. 

We have calculated $\Delta_2(N)=E_{N+1}+E_{N-1}-2E_{N}$
in the GSE and GUSE cases and compared with the
experimental data of S.R. Patel et al., \cite{Siv1}.
Here $E_{N}$ is the ground state energy of QD with $N$ electrons. 
The results are shown in Fig.1. 
Our calculations in obtaining these distributions were kept at a very simple level.
The GUSE was the simplest case since it did not have non trivial interaction
terms in the leading 1/M order, Eq. (\ref{9b}). We used the HF expressions 
for $E(N)$ and obtained
\begin{eqnarray}  \label{10}
\Delta_2^{GUSE}(N)=V_c+\epsilon_{N+1}-\epsilon_{N}+ u^n_{N+1,N+1}+u^n_{N,N}
\end{eqnarray} 
  
\begin{figure}
\centering
\epsfig{figure=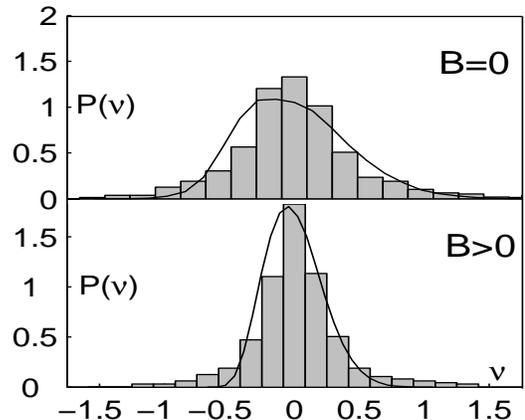,height=6.cm,width=7.cm}
\caption{$\nu=(\Delta_2-<\Delta_2>)/\Delta$ - normalized peak spacings 
for $B=0$ and $B\ne 0$. Histograms are experimental data,
solid lines  - predictions of the interacting GSE (GUSE) for $B=0$ ($B\ne 0$)}
\label{fig1} \end{figure}
 We used RMT GUE distribution for $\epsilon_{N+1}-\epsilon_{N}$ 
and the Gaussian distribution for  $u^n_{N+1,N+1}$, $u^n_{N,N}$ 
with the covariance $C_{nn}=0.069$. This value as well as $C_{np}$ and $C_{pp}$
below were obtained using a reasonable parametrization of the screened
e-e interaction, \cite{BGL1}. 

For GSE the calculations required a proper treatment of the pairing
$PT^{\dagger}T$ interaction appearing in the leading order in (\ref{9a}).
This problem has an exact solution \cite{Rich}. However
we used a simple approximation which we felt was satisfactory. For even N we
minimized this term in the subspace of two HF solutions with  adjacent
filled and empty Kramers pairs. We then formed an expectation of H with the
resulting wave function. For the odd N the effect of the pairing is much simpler
and the expectation with lowest energy HF wave function was
sufficient. The resulting expressions are too cumbersome to record, cf.,
\cite{BGL1}. We used them with the RMT GSE statistics, $P=0.37\Delta$,
the covariance $C_{nn}$ as in GUSE and $C_{pp}=0.023$, $C_{np}=0$. 

\begin{figure}
\centering
\epsfig{figure=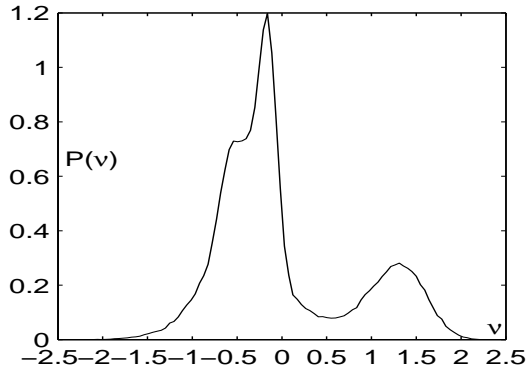,height=5.cm,width=7.cm}
\caption{Normalized peak spacings  for the interacting GUE}
\label{fig2} \end{figure}

As one can see in Fig. 1 there no sign of the bimodal structure in the GSE 
and GUSE distributions. The reason for this in GUSE  is perfectly obvious, cf.,
Eq. (\ref{10}) - the spin degeneracy which is responsible for the
bimodal structure in simple  models of the CB is completely
washed out by the combined effect of the SO interaction and the magnetic field.
In the interacting GSE the non commutativity of the 
paring term $PT^{\dagger}T$ with
the single particle part causes rather strong fluctuations 
relative to the RMT already in the lowest order. To appreciate this effect it
is instructive to 
compare the leading GSE interaction terms, upper line in (\ref{9a}) with those 
of GUE, the upper line with $P=0$ in (\ref{9}). The commuting spin interaction 
$J\hat{\bf S}^2$ does not change the basic RMT fluctuations but simply cuts 
and shifts different spin parts creating sharp structures. As seen in Fig. 2
these structures are  washed out only partially by higher order terms. 
We conclude by observing that our results, Fig.1,  fit quite poorly the
tails of the spacing distributions. It is not clear to us if this is a 
consequence of our approximations in calculating $\Delta_2(N)$ or because of 
more fundamental reasons. 

We wish to express our thanks to B.L. Altshuler, 
A. Finkelstein, K. Kikoin, Y.Oreg and J. da
Providencia for useful discussions. V.B. and S.L. express their thanks for hospitality of 
University of Coimbra  where part of this work was done. 
This work was supported in part by the DIP grant DIP-C 7.1. V.B. was supported
in part by the NATO Science Fellowship Program CPRU18C00P0.

\widetext
\end{document}